\documentclass[aps,prl,preprint]{revtex4}

\usepackage{graphicx}
\usepackage{amsmath,amssymb}

\begin{document}

\title{Phaseless three-dimensional optical nano-imaging}

\author{Alexander A. Govyadinov}
\email{algov@seas.upenn.edu}
\author{George Y. Panasyuk}
\email{georgey@seas.upenn.edu}
\author{John C. Schotland}
\email{schotland@seas.upenn.edu}

\affiliation{Department of Bioengineering, University of Pennsylvania, Philadelphia, Pennsylvania}

\date{\today}

\begin{abstract}
We propose a method for optical nano-imaging in which the structure of a three-dimensional inhomogeneous medium may be recovered from far-field power measurements.  Neither phase control of the illuminating field nor phase measurements of the scattered field are necessary. The method is based on the solution to the inverse scattering problem for a system consisting of a weakly-scattering dielectric sample and a strongly-scattering nano-particle tip. Numerical simulations are used to illustrate the results.

\end{abstract}


\maketitle

The development of tools for three-dimensional imaging of nanostructures is of considerable current interest~\cite{Taubner2005a,Anderson2006,Cvitkovic_2007,Aizpural_2008}. There are multiple potential applications, including inspection of semiconductor devices, detection of atoms buried beneath surfaces and characterization of biologically important supramolecular assemblies, among others. Optical methods, especially near-field scanning optical microscopy (NSOM) and its variants, hold great promise for nanoscale imaging due to their subwavelength resolution,  spectroscopic sensitivity to chemical composition and nondestructive nature~\cite{Novotnybook}. 
Although traditionally viewed as a technique for imaging surfaces, near-field microscopy has recently demonstrated the capacity to detect subsurface structure~\cite{Taubner2005a,Cvitkovic_2007}. Experiments in which a near-field probe is scanned over a three-dimensional volume outside the sample suggest that information on the three-dimensional structure of the sample is encoded in the data. That is, the measured intensity viewed as a function of height above the sample is seen to depend upon the depth of subsurface features. However, the intensity images obtained in this manner are not tomographic, nor are they quantitatively related to the optical properties of the medium.

The above noted difficulties have led to the use of inverse-scattering theory to elucidate the precise manner in which three-dimensional subwavelength structure
is encoded in the optical near-field~\cite{Carminati,Carney2000,CarneySchotlandOL01,CarneyMarkelSchotlandPRL01,Carney2004,Chaumet2004b,Belkebir2005,Sentenac2006,Panasyuk2006,Li2007,Sun_2007,Baleine_2005}. Results in this direction have been reported for two-dimensional reconstructions of thin samples~\cite{Carney2004} and also for three-dimensional inhomogeneous media~\cite{Gaikovich}. In either case, solution of the inverse scattering problem generally requires measurements of the optical phase, in the form of a near-field hologram, an experiment that is notorious for its difficulty. The replacement of  phase measurements by phase control of the illuminating field has also been proposed~\cite{CarneyMarkelSchotlandPRL01}. In this approach, the power extinguished from an incident evanescent wave field is used to reconstruct the imaginary (absorptive) part of the dielectric susceptibility, leaving the real part unrecoverable. 

In this Letter we propose a method for nano-scale optical tomography
that relies neither on phase-measurements of the scattered field nor on phase-control of the illuminating field. Our approach enables the reconstruction of the \emph{complex-valued} dielectric susceptibility with subwavelength resolution in three dimensions. As a proxy for the optical phase, we introduce a controlled scatterer, such as an atomic force microscopy (AFM) tip, into the near-field of the sample. The power extinguished from the incident field, which illuminates the sample and the tip, is then measured in the {\emph {far-field}}. Since the tip is placed externally  to the sample, changing its position controls the pattern of illumination, which thus modifies the power extinguished from the incident beam. The crucial difference from conventional holographic techniques is that the interference pattern is regulated by an \emph{internal} degree of freedom of the system (regarded as the sample plus the tip) rather than by external illumination. The burden of phase-resolved measurements or phase-controlled illumination is thus replaced by the problem of controlling the position of the tip. The readily available nanometer precision in probe positioning that is achievable in AFM, in combination with the simplicity of far-field measurements of the extinguished power, is expected to allow the practical realization of the proposed method.

\begin{figure}[t] 
\includegraphics[width=6cm]{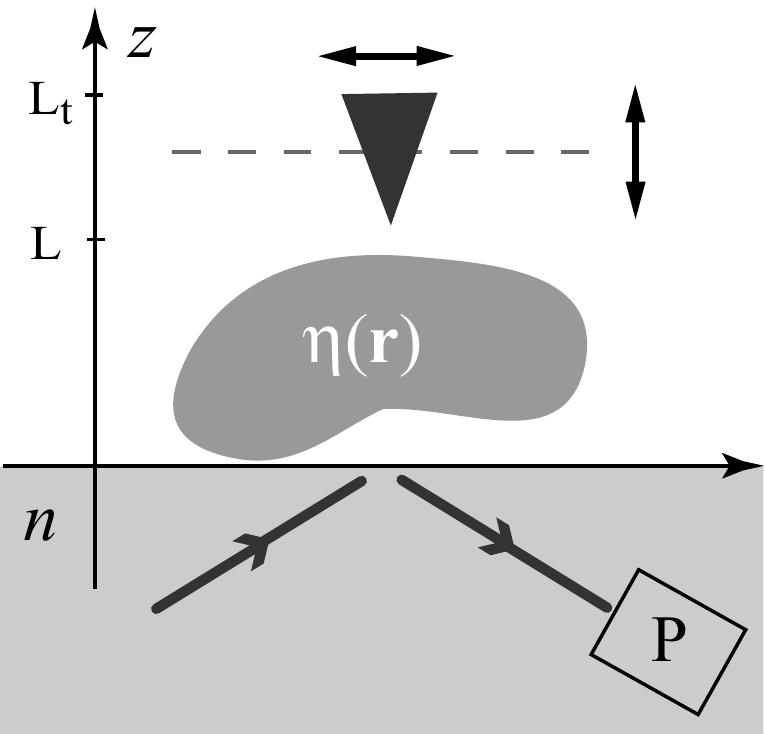}
\caption{\label{fig:schematic} Illustrating the experiment. The tip scatters the incident evanescent field and modifies the interference pattern in the sample, which has dielectric susceptibility $\eta(\bf r)$. The power $P$ extinguished from the illuminating field is measured as the tip is scanned on a three-dimensional grid in the near-zone of the sample.}
\end{figure}

We begin by considering an experiment in which a sample is deposited on a planar substrate. The lower half-space $z<0$ (the substrate) is taken to have a constant index of refraction $n$. The sample occupies the upper half-space $z\ge 0$ and is assumed to be nonmagnetic. The index of refraction in the upper half-space varies within the sample, but otherwise has a value of unity. The upper half-space also contains the tip which is placed in the near-field of the sample. The sample and tip are illuminated from below by a monochromatic evanescent plane wave and the power extinguished from the illuminating field is monitored, as shown in Fig.~\ref{fig:schematic}. 

The electric field $\mathbf{E}$ in the upper half-space obeys the reduced wave equation
\begin{equation}
\label{eq:wave}
\nabla\times\nabla\times\mathbf{E}(\mathbf{r}) - k_0^2 \mathbf{E}(\mathbf{r}) = 4\pi k_0^2 \left(\eta(\mathbf{r})+\chi(\mathbf{r})\right) \mathbf{E}(\mathbf{r}) \ ,
\end{equation}
where $\eta$ is the generally-complex dielectric susceptibility of the sample, $\chi$ is the susceptibility of the tip, $k_0=2\pi/\lambda$ is the free-space wave number and the field obeys the necessary interface and boundary conditions. The field is taken to consist of two parts, $\mathbf{E}=\mathbf{E}_i+\mathbf{E}_s$, where $\mathbf{E}_i$ is the incident field and $\mathbf{E}_s$ is the scattered field. The incident field obeys Eq.~(\ref{eq:wave}) in the absence of the sample and the tip. The scattered field obeys the integral equation
\begin{equation}
\label{eq:UsGeneral}
{\bf E}_s(\mathbf{r})=k_0^2\int {\bf \bar G}(\mathbf{r},\mathbf{r}^\prime) \cdot {\bf E}(\mathbf{r}^\prime) \left(\eta(\mathbf{r}^\prime)+\chi(\mathbf{r}^\prime)\right) d^3r^\prime \ ,
\end{equation}
where ${\bf \bar G}$ is the half-space Green's tensor. The power $P_e$ extinguished from the illuminating field can be obtained using the generalized optical theorem~\cite{OptTh}:
\begin{equation}
\label{eq:OptTh}
P_e=\frac{k_0c}{2}{\rm Im} \int_V \mathbf{E}_{i}^*(\mathbf{r})\cdot\mathbf{E}(\mathbf{r})\left(\eta(\mathbf{r})+\chi(\mathbf{r})\right)d^3r \ ,
\end{equation}
where the integration is performed over the volume $V$, which contains both the sample and the tip. 

Suppose that the tip is a strongly-scattering, possibly metallic, nanoparticle and the sample is a weakly-scattering dielectric. We may then compute the electric field perturbatively, accounting for all orders of scattering from the tip and one order of scattering from the sample. We find that the resulting perturbation series can be resummed and, neglecting contributions arising solely from the sample or the tip, consists of a sum of three terms~\cite{Sun_2007}. The first, or `TS', term corresponds to scattering from the tip and then from the sample. The second, or `ST', term is due to scattering from the sample and then from the tip. The third, or `TST' term, arises from scattering first from the tip, then from the sample and finally from the tip. Note that two additional terms originating solely from scattering by the sample or the tip contain no structural information and will be omitted. In practice they can be removed by calibration. To proceed further, we must specify a model for the tip. We treat the tip as a small scatterer with susceptibility $\chi(\mathbf{r})=\alpha_0\delta(\mathbf{r}-\mathbf{r}_t)$, where $\mathbf{r}_t$ is the tip's position and $\alpha_0$ is its polarizability.
Resummation of the perturbation series, as explained above, leads to a renormalization of the polarizability of the tip of the form $\alpha=\alpha_0/(1-2ik^3\alpha_{0}/3)$, which includes the lowest order radiative corrections to the bare polarizability but neglects the dependence on the tip height above the interface~\cite{deVries_1998,Sun_2007}.

It follows from Eq.~(\ref{eq:OptTh}) that the extinguished power can be expressed as a sum of contributions of ST, TS and TST types:
\begin{align}
\label{eq:Pintegral}
	P_e(\mathbf{r}_t)=\frac{ck_0^3}{4i}\int \sum_{p=1}^2K^{(p)}(\mathbf{r}_t,\mathbf{r})\eta^{(p)}(\mathbf{r}) d^3r \ ,
\end{align}
where $\eta^{(1)}(\mathbf{r})={\eta^{(2)*}}(\mathbf{r})=\eta(\mathbf{r})$, the kernels $K^{(p)}(\mathbf{r}_t,\mathbf{r})$ are defined by
\begin{widetext}
\begin{align}
\label{eq:kernel}
\nonumber
K^{(1)}(\mathbf{r}_t,\mathbf{r}) =-{K^{(2)}}^*(\mathbf{r}_t,\mathbf{r})= \alpha \mathbf{E}_i^*(\mathbf{r}_t) \cdot {\bf \bar G}(\mathbf{r}_t,\mathbf{r}) \cdot \mathbf{E}_i(\mathbf{r}) + \alpha \mathbf{E}_i^*(\mathbf{r})\cdot {\bf \bar G}(\mathbf{r},\mathbf{r}_t) \cdot \mathbf{E}_i(\mathbf{r}_t) \\ + 
\alpha^2 k^2 \mathbf{E}_i^*(\mathbf{r}_t)\cdot\left( {\bf \bar G}(\mathbf{r}_t,\mathbf{r}) {\bf \bar G}(\mathbf{r},\mathbf{r}_t)\right) \cdot \mathbf{E}_i(\mathbf{r}_t)
 \end{align}
\end{widetext}
and the dependence of the extinguished power on the tip position has been made explicit.

We will assume that the sample occupies the region $0\le z \le L$ and that it is illuminated by a plane wave of the form $\mathbf{E}_i(\mathbf{r})=\mathbf{E}_0 \exp(i \mathbf{q}_i\cdot{\boldsymbol \rho}+{k}_z z)$. Here $ \mathbf{r}=(\boldsymbol\rho,z)$ and the field has amplitude $\mathbf{E}_0$, transverse wave vector $\mathbf{q}_i$, and ${k}_z=\sqrt{(nk_0)^2-q_i^2}$. The extinguished power is measured for a discrete set of tip positions located on a three-dimensional cartesian grid with transverse spacing $h$ and longitudinal spacing $\Delta z$. Note that the tip occupies the region $L < z \le L_t$ and thus does not overlap the sample.

It will prove useful to perform a two-dimensional lattice Fourier transform of 
the sampled extinguished power in the plane $z=z_t$, namely $\tilde{P_e}(\mathbf{q},z_t)=\sum_{\boldsymbol \rho}\exp(i\mathbf{q}\cdot\boldsymbol \rho)P_e(\boldsymbol\rho,z_t)$. Here the sum is carried out over all lattice vectors and $\mathbf{q}$ is restricted to the first Brillouin zone (FBZ) of the lattice. Next, we require the plane-wave decomposition of the tensor Green's function
\begin{equation}
{\bf \bar G}(\mathbf{r,r'})=\int \frac{d^2q}{(2\pi)^2} \exp[i \mathbf{q}\cdot({\boldsymbol \rho}-{\boldsymbol \rho^\prime})]{\bf \bar g}_{\bf q}(z,z^\prime) \ ,
\end{equation}
where the form of ${\bf \bar g}_{\bf q}$ is given in Ref.~\cite{Maradudin}. Making use of this result and carrying out the lattice Fourier transform, we find that Eq.~\eqref{eq:Pintegral} becomes
\begin{align}
\label{eq:PFT}
\tilde{P_e}(\mathbf{q},z_t)=\int_0^L  \sum_{p=1}^2 \tilde{K}^{(p)}(\mathbf{q};z_t,z)\tilde{\eta}^{(p)}(\mathbf{q},z)dz \ ,
\end{align}
where $\tilde{K}^{(1)}(\mathbf{q};z_t,z)$ is defined as
\begin{widetext}
\begin{align}
\label{eq:kernelFT}
\nonumber
\tilde{K}^{(1)}(\mathbf{q};z_t,z)=\alpha T \Bigg(\gamma\mathbf{E}_0^*\cdot {\bf \bar g}_{\mathbf{q}_i-\mathbf{q}}(z_t,z)\cdot \mathbf{E}_0
+\gamma^*\mathbf{E}_0^* \cdot {\bf \bar g}_{\mathbf{q}_i+\mathbf{q}}(z,z_t)\cdot\mathbf{E}_0 \\ 
+\alpha k_0^2\int\frac{d^2q^\prime}{(2\pi)^2}
\mathbf{E}_0^* \cdot\left({\bf \bar g}_{\mathbf{q^\prime}}(z_t,z) {\bf \bar g}_{\mathbf{q}+\mathbf{q}^\prime}(z,z_t)\right)\cdot\mathbf{E}_0\Bigg) 
\end{align}
\end{widetext}
and $\tilde K^{(2)}({\bf q};z_t,z)= \tilde K^{(1)*}(-{\bf q};z_t,z)$. Here $T=ck_0^3/(4ih^2)\exp[-2{\rm Im}{k}_z z_t]$, $\gamma=\exp[i {k}_z (z-z_t)]$ and $\tilde{\eta}(\mathbf{q},z)=\int d^2\rho \exp(i \mathbf{q} \cdot {\boldsymbol \rho}) \eta(\boldsymbol\rho,z)$. Note that for fixed $\bf q$, Eq.~\eqref{eq:PFT} defines a one-dimensional integral equation for $\tilde\eta^{(p)}({\bf q},z)$. 

The inverse scattering problem we consider consists of recovering $\eta^{(p)}$, for $p=1,2$, from measurements of $P_e$. This corresponds to solving the integral equation Eq.~\eqref{eq:PFT}. If it is known, apriori, that the susceptibility $\eta$ is purely real or imaginary, then the inverse problem is formally determined and the solution to Eq.~\eqref{eq:PFT} is readily obtained by singular value decomposition (SVD)~\cite{Carney2000,CarneyMarkelSchotlandPRL01}. However, if $\eta$ is complex-valued, the inverse problem is underdetermined. To resolve this difficulty, it is necessary to introduce additional data, which we take to consist of a second set of measurements. That is, two sets of measurements must be carried out for each location of the tip, yielding $\tilde{P_e}_1$ and $\tilde{P_e}_2$, one for each incident plane wave with transverse wave vectors $\mathbf{q}_{1,2}$ and amplitudes $\mathbf{E}_{1,2}$, respectively. In this manner, it is possible to reconstruct both $\eta$ and $\eta^*$ simultaneously, which is equivalent to recovering the real and imaginary parts of $\eta$ from power extinction measurements. Following the approach of Ref.~\cite{Markel_2003}, we find that the solution to the integral equation~\eqref{eq:PFT} is given by the formula
\begin{eqnarray}
\label{eq:eta}
\nonumber
\tilde{\eta}^{(p)}(\mathbf{q},z) = \sum_{z_t,z_t^\prime}
\sum_{i,j} {\tilde{K}_i^{(p)*}} (\mathbf{q};z_t,z)M_{ij}^{-1}(\mathbf{q};z_t,z_t^\prime) \tilde{P}_{ej}(\mathbf{q},z_t^\prime) \ ,
\end{eqnarray}
where $i,j=1,2$ label the incident waves.
Here $M_{ij}^{-1}$ is the inverse of the matrix whose elements are given by
\begin{equation}
\label{eq:M} 
M_{ij}(\mathbf{q};z_t,z_t^\prime)=\int_0^L
\sum_{p} \tilde{K}_i^{(p)}(\mathbf{q};z_t,z)\tilde{K}^{(p)*}_j (\mathbf{q};z_t^\prime,z) dz \ .
\end{equation}
An inverse Fourier transform is then applied to obtain
a transversely bandlimited approximation to $\eta(\bf r)$ with bandwidth $2\pi/h$. 

\begin{figure}[t]
\includegraphics[width=8.5cm]{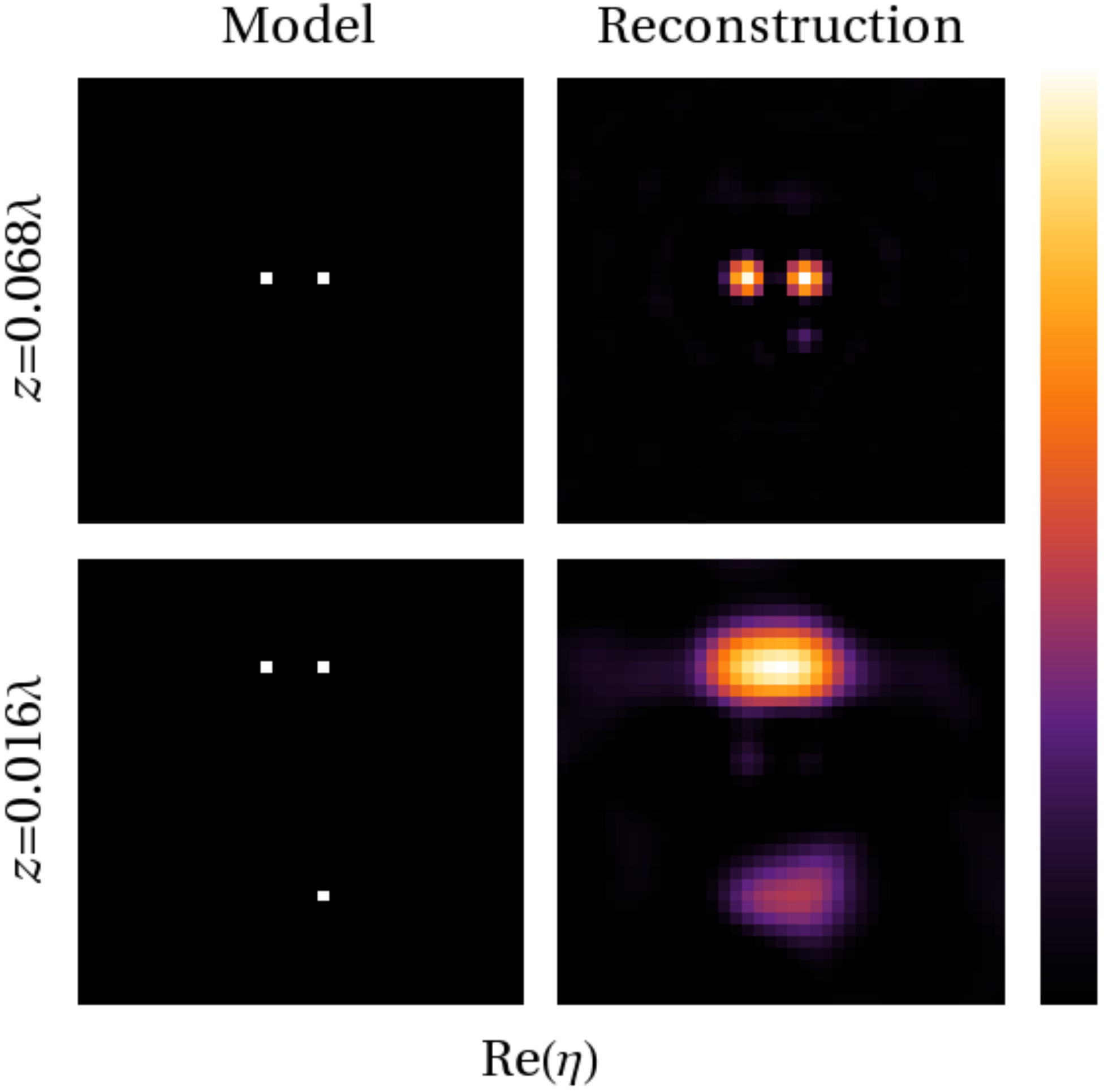}
\includegraphics[width=8.5cm]{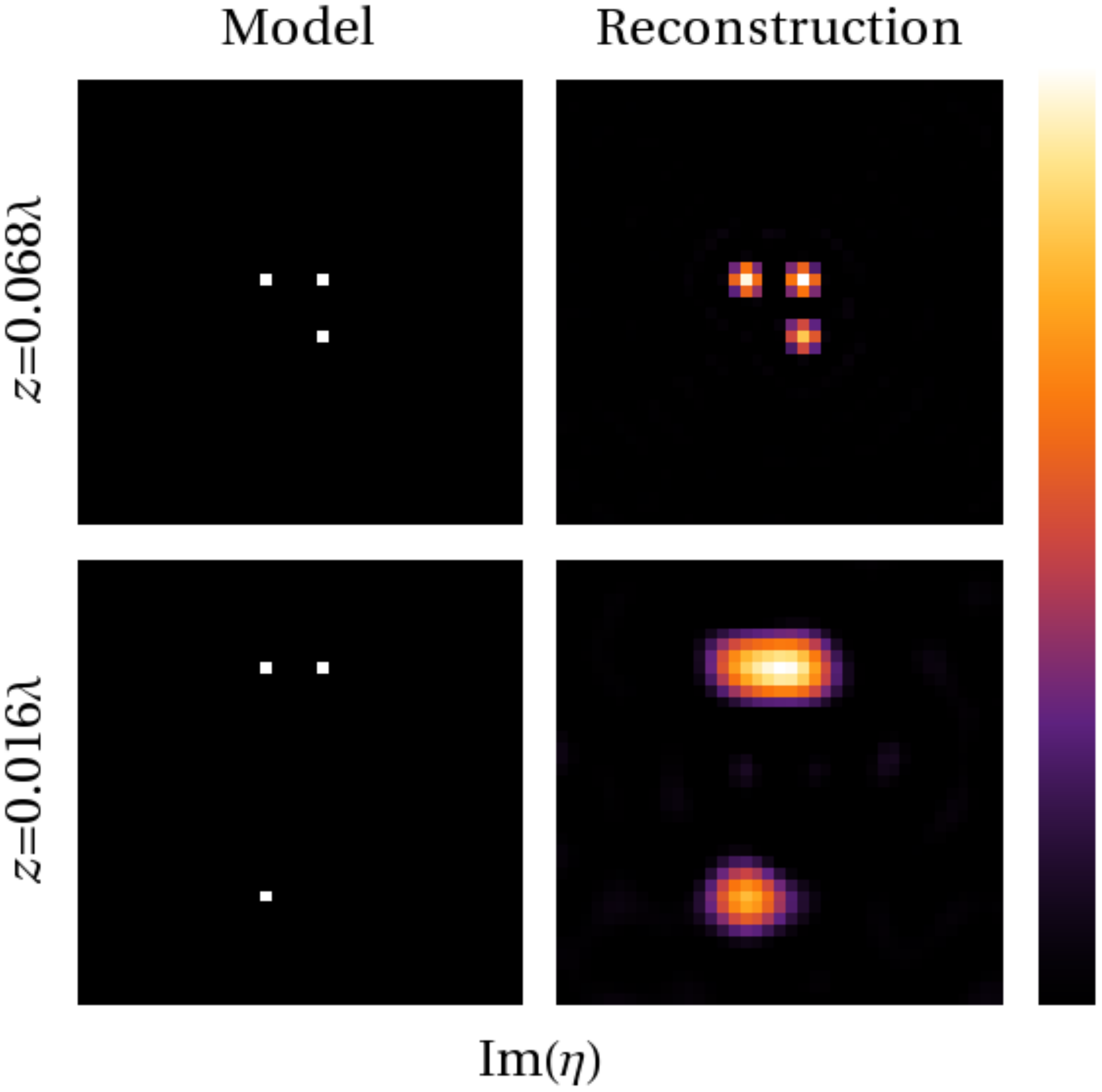}
\caption{\label{fig:recon}(color online) The model (left) and simulated reconstructions (right) of ${\rm Re}\eta(\mathbf{r})$ and ${\rm Im}\eta(\mathbf{r})$. The scatterers are distributed in two planes at $z=0.016\lambda$ (top) and $z=0.068\lambda$ (bottom). The scatterers are separated in-plane by $0.05\lambda$ and $0.2\lambda$ in the $x$- and $y$-directions, respectively. Each image is normalized to its own maximum and any small negative values are not displayed. The field of view of each image is $0.4\lambda \times 0.4 \lambda$.}
\end{figure}

To demonstrate the feasibility of the inversion, we have numerically simulated the reconstruction of $\eta(\mathbf{r})$ for a collection of point scatterers. The left column in Fig.~\ref{fig:recon} shows the configuration of the scatterers. The tip was modeled as a small sphere of polarizability $\alpha_0 = (\epsilon-1)/(\epsilon+2)R^3$ with radius $R=8\times10^{-2}\lambda$ and permittivity $\epsilon=-11.39+0.13i$, which corresponds to silver at a wavelength $\lambda=550$ nm. The incident fields were taken to be evanescent plane waves with transverse wave vectors $\mathbf{q}_1=(3.15k_0/\pi,0)$ and $\mathbf{q}_2=(0,3.25k_0/\pi)$, and vector amplitudes $\mathbf{E}_1=(-0.521,-0.714,0.468)$ and $\mathbf{E}_2=(-0.507,-0.714,0.483)$, respectively.  The susceptibility $\eta$ was reconstructed on a $40\times 40\times 20$ cartesian grid whose transverse extent was $0.4\lambda \times 0.4 \lambda$ and height in the $z$-direction was $0.08 \lambda$. The forward data were calculated from Eq.~\eqref{eq:Pintegral} for the positions of the tip center located on the same $40\times 40$ transverse grid with $20$ steps of size $\Delta z = 0.001\lambda$ in the $z$ direction,  beginning $0.16\lambda$ from substrate. The integral in the kernel \eqref{eq:kernelFT} was numerically evaluated using a trapezoidal rule with $300$ points in each direction, spanning six Brillouin zones. We note that computation of the matrix inverse $M_{ij}^{-1}$ requires regularization, which we 
carry out by retaining only those singular values in the SVD of $M_{ij}$ which are larger than a prescribed cutoff.

In Fig.~\ref{fig:recon} we present reconstructions of the real and imaginary parts of $\eta$. Tomographic slices are shown in the planes $z=0.016\lambda$ and $z=0.068\lambda$. It can be seen that the scatterers in the top layer (nearest the tip) are better resolved than the scatterers in the deeper layer. This is due to the decay of high-frequency evanescent waves with depth and is a typical feature of tomographic reconstructions in the near-field~\cite{Carney2000}. It may also be observed that the reconstructions of the imaginary part of the susceptibility are of higher quality than those of the real part. This effect may be explained by noting that the extinction of power due to absorption is greater than that due to elastic scattering in the near-field, since the optical phase changes minimally in the near-zone of the scatterer.

In conclusion, we have shown that the three-dimensional \emph{subwavelength} structure of an inhomogeneous scattering medium can be recovered from \emph{far-field} measurements of the extinguished power. Remarkably, neither phase control of the illuminating field nor phase measurements of the scattered field are required. Our approach is based on the solution to the inverse scattering problem for a system consisting of a weakly-scattering sample and a strongly scattering nano-scale tip. In principle, the theory can be extended to treat the case of a strongly-scattering sample by inversion of an appropriately  resummed perturbation series, taken to all orders of scattering in the sample~\cite{Panasyuk2006}. 
It is important to note that the observed subwavelength resolution is due to the modification of the near-field of the sample, resulting from the presence of the tip. It is not due to the illuminating evanescent wave, as is the case in total internal reflection tomography~\cite{CarneySchotlandOL01,Chaumet2004b}. The evanescent wave is introduced solely for the experimental convenience of measuring the extinguished power.

The authors are grateful to Prof.~Vadim A. Markel for valuable discussions.
This work was supported by the NSF under the grant DMR0425780 and by 
the USAFOSR under the grant FA9550-07-1-0096.

\newpage

\end{document}